\documentclass[prl,a4paper,twocolumn,superscriptaddress,longbibliography]{revtex4-1}

\usepackage{amssymb}
\usepackage{amsmath}
\usepackage{amsfonts}
\usepackage{graphicx}
\usepackage{bm}
\usepackage{xcolor}
\usepackage{multirow}
\usepackage{natbib}
\usepackage{hyperref}
\usepackage[normalem]{ulem}
\usepackage{relsize}
\usepackage[final]{pdfpages}
\usepackage{pgffor}

\makeatletter
\AtBeginDocument{\let\LS@rot\@undefined}
\makeatother

\begin{document}

\title{The effect of anomalous elasticity on the bubbles in van der Waals heterostructures}

\author{A. A. Lyublinskaya}

\author{S. S. Babkin} 

\affiliation{Moscow Institute of Physics and Technology, 141700, Dolgoprudnyi, Moscow Region, Russia}


\author{I. S. Burmistrov}

\affiliation{L. D. Landau Institute for Theoretical Physics, Semenova 1-a, 142432, Chernogolovka, Russia}
\affiliation{Laboratory for Condensed Matter Physics, National Research University Higher School of Economics, 101000 Moscow, Russia}

\begin{abstract}
It is shown that the anomalous elasticity of membranes affects the profile and thermodynamics of a bubble in van der Waals heterostructures. Our theory generalizes the non-linear plate theory as well as membrane theory of the pressurised blister test to incorporate the power-law scale dependence of the bending rigidity and Young's modulus of a two-dimensional crystalline membrane. This scale dependence caused by long-ranged interaction of relevant thermal fluctuations (flexural phonons), is responsible for the anomalous Hooke's law observed recently in graphene. It is shown that this anomalous elasticity affects dependence of the maximal height of the bubble on its radius and temperature. We identify the characteristic temperature above which the anomalous elasticity is important. It is suggested that for graphene-based van der Waals heterostructures the predicted anomalous regime is experimentally accessible at the room temperature.
\end{abstract}

\maketitle

\textsf{Introduction.}\ --- Mechanical properties of two-dimensional  (2D) materials, especially, of van der Waals heterostructures, have recently attracted a lot of interest  in view of their potential applications \cite{Dai2019}. The simplest example of van der Waals heterostructure is two monolayers, e.g.
 graphene, hexagonal boron nitride (hBN), MoS$_2$, assembled together. Strong adhesion between monolayers \cite{Megra2019} results in atomically clean interfaces in which all contaminating substances are combined into bubbles \cite{Haigh2012}. Recently, these bubbles inside van der Waals heterostructures have been studied experimentally \cite{Khestanova2016}. Similar bubbles are formed between an atomic monolayer and a substrate, e.g. SiO$_2$ \cite{Khestanova2016,Dai2018}. There are many suggestions of  practical usage of the bubbles inside van der Waals heterostructures, for example, the 
graphene liquid cell microscopy \cite{Ghodsi2019}, 
controlled room-temperature photoluminescence emitters \cite{Tyurnina2019}, etc.

The mechanics of monolayers due to these bubbles is considered to be analogous to the one of the pressurized blister test  which has recently become the routine method to measure simultaneously the Young's modulus and adhesion energy of a monolayer on a substrate \cite{Bunch2011,Boddeti2013,Lloyd2017,Wang2019}. Usually, the pressurized blister test is described either by  nonlinear plate model or by membrane theory (see e.g. \cite{Yue2012,Wang2013}). These standard elastic theories of deformed plates ignore the fact that elastic properties of an atomic monolayer are those of 2D crystalline membranes \cite{Nelson1987,Aronovitz1988,Paczuski1988,David1988,Aronovitz1989,Guitter1989,Doussal1992}. The most striking feature of mechanics of membranes is anomalous elasticity which results in non-linear (so-called, anomalous) Hooke's law for small applied stress (see Refs. \cite{BookNelson,Doussal2018} for a review).
Recently, this anomalous Hooke's law has been measured in graphene \cite{Nicholl2015,Gornyi2016}. However, until present, see e.g. Refs. \cite{Ghorbanfekr2017,Zhang2017,Delfanu2018,Sanchez2018}, the anomalous elasticity of membranes is completely ignored in description of mechanical properties of monolayers in the presence of the bubbles.

The present paper generalizes the classical theory of the pressurised blister test to incorporate the anomalous elasticity of a membrane. Our approach explicitly takes into account the power-law renormalization of the bending rigidity and Young's modulus. It is shown that above a certain temperature the dependence 
of the bending rigidity and Young's modulus of a membrane on the radius  of the bubble results in the non-analytic dependence of its maximal height on the radius and temperature.

\begin{figure}
\includegraphics[width=.8\linewidth]{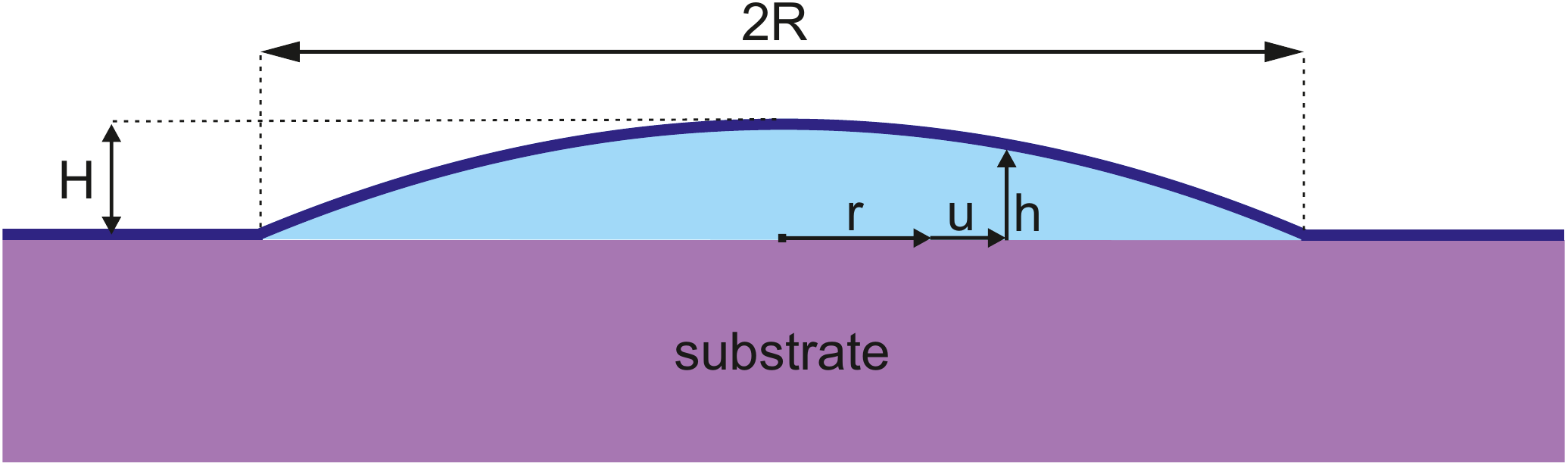}
\caption{Sketch of a spherical bubble between a membrane and a substrate.}
\label{Figure:Bubble}
\end{figure}

\textsf{Model.}\ --- The model for the description of the profile of a bubble between the membrane  and the substrate (see Fig. \ref{Figure:Bubble}) is well established \cite{Yue2012,Khestanova2016}. It can be formulated in terms of the (free) energy which is the sum of the following four terms:    
\begin{equation}
E=E_{\rm bend}+E_{\rm el}+E_{\rm b}+E_{\rm vdW} .
\label{eq:energy}
\end{equation}
Here the first contribution describes the energy cost related with bending of the membrane:
\begin{equation}
E_{\rm bend} = \frac{\varkappa_0}{2}   \int d^2 \bm{r} \bigl[ (\Delta h)^2 + (\Delta \bm{u})^2\Bigr ] ,
\label{eq:bending:energy}
\end{equation}
where $\varkappa_0$ denotes the bare bending rigidity of the membrane.  Here $\bm{u}=\{u_x,u_y\}$ and $h$ are the in-plane and out-of-plane displacements of the membrane (see Fig. \ref{Figure:Bubble}). The second term is the standard elastic energy \cite{Landau}:
\begin{equation}
E_{\rm el} = \int d^2 \bm{r} \bigl (
\mu_0 {u}_{\alpha\beta}  {u}_{\beta\alpha}
+\lambda_0 {u}_{\alpha\alpha} {u}_{\beta
\beta} /2 \bigr ) ,
\label{eq:elastic:energy}
\end{equation}
where $\mu_0$ and $\lambda_0$ stand for the Lam\'e coefficients and  $
{u}_{\alpha\beta} = \bigl ( \partial_\beta u_\alpha + \partial_\alpha u_\beta  + \partial_\alpha h \partial_\beta h + \partial_\alpha \bm{u}  \partial_\beta \bm{u}\bigr )/2$ is the strain tensor.
The  third term $E_{\rm b}$ describes the contribution of the bubble substance. Under assumption of the constant pressure $P$ inside the bubble one has $E_{\rm b} = - P V$, where the bubble volume can be approximated as 
$V = \int d^2 \bm{r}\,  h(\bm{r})$. The last term in Eq. \eqref{eq:energy} describes the van der Waals interaction between the membrane and the substrate in the presence of the bubble. It can be approximated as
$E_{\rm vdW} = \pi \gamma R^2$.  Here $R$ denotes the radius of the bubble (see Fig. \ref{Figure:Bubble}) and  $\gamma= \gamma_{\rm ms}-\gamma_{\rm mb}-\gamma_{\rm sb}$ where $\gamma_{\rm ms}$, $\gamma_{\rm mb}$, and $\gamma_{\rm sb}$
are the adhesion energies between the membrane and substrate, between the membrane and the substance inside the bubble, and between the substrate and the substance, respectively. The form \eqref{eq:energy} of the total energy is well justified if the maximal height of the bubble, $H=h(0)$, is small compared to the radius, 
$H \ll R$.
Throughout the paper we shall assume that this condition is fulfilled.

\textsf{The standard approach.}\ --- In order to compute the profile of the spherical bubble, initially, one needs to solve the Euler-Lagrange equations for $\bm{u}(r)$ and $h(r)$ with the proper boundary conditions and compute the energy $E$ as a function of $R$ and $H$. Usually, instead of the solution of the Euler-Lagrange equations the approximate solutions either within the non-linear plate theory or within the membrane theory are used, see e.g. Ref. \cite{Yue2012}. Finally, one have to minimize $E$ with respect to the both $R$ and $H$. The minimization procedure allows one to find the maximal height $H$ and the pressure $P$ as a function of the bubble radius $R$. 
Comparison of the linear in $u$ term and the term quadratic in $h$ in the strain tensor $u_{\alpha\beta}$ leads to the following relation for the maximal horizontal deformation: $u_{\rm max} \sim H^2/R$. In the considered regime, $H/R \ll 1$,
 the horizontal displacement is also small, $u_{\rm max} \ll H$. This implies that one can neglect the term $\partial_\alpha \bm{u}  \partial_\beta \bm{u}$ in $u_{\alpha\beta}$
. Also this allows one to omit the term $(\Delta \bm{u})^2$ in the bending energy such that it reads:
\begin{equation}
E_{\rm bend} = \frac{\varkappa_0}{2}   \int d^2 \bm{r} (\Delta h)^2 .
\label{eq:bending:energy:mod}
\end{equation}

In the absence of $\partial_\alpha \bm{u}  \partial_\beta \bm{u}$ in $u_{\alpha\beta}$ the elastic energy becomes quadratic in $\bm{u}$. This implies that the Euler-Lagrange equation for $\bm{u}(\bm{r})$ become linear and can be solved for arbitrary configuration of $h(\bm{r})$ (even not necessarily obeying the Euler-Lagrange equation). In other words, a horizontal deformation is adjusted to any vertical displacement. Therefore, $E_{\rm el}$ is given as \cite{Nelson1987}:
\begin{equation}
E_{\rm el} = \frac{Y_0}{8}\! \int \!d^2\bm{r}  \left [K_{\alpha\alpha}
-
\partial_\alpha\!
\int d^2\bm{r^\prime} 
\mathcal{G}(\bm{r},\bm{r^\prime})
\partial_\beta
K_{\alpha\beta}(\bm{r^\prime}) \right ]^2 ,
\label{eq:elastic:energy:mod}
\end{equation}
where $K_{\alpha\beta} = \partial_\alpha h \partial_\beta h$ and $Y_0 = \frac{4\mu_0(\mu_0+\lambda_0)}{2\mu_0+\lambda_0}$ is the Young's modulus. The function $\mathcal{G}(\bm{r},\bm{r^\prime})$ is the Green's function of the Laplace operator on the disk $r\leqslant R$.

Using Eqs. \eqref{eq:bending:energy:mod} and \eqref{eq:elastic:energy:mod}, we can estimate the bending and elastic energies as 
$E_{\rm bend} \sim \varkappa_0 H^2/R^2$ and
$E_{\rm el} \sim Y_0 H^4/R^2$. For $H \gg a$, where $a \sim \sqrt{\varkappa_0/Y_0}$ is the effective thickness of the membrane, the elastic energy dominates over bending energy, $E_{\rm el}\ll E_{\rm bend}$. We note that typically, the effective thickness is smaller than the lattice spacing, e.g.  for graphene $a \sim 1\, \AA$. Thus, by neglecting $E_{\rm bend}$ and minimizing $E_{\rm el}+ E_{\rm b} + E_{\rm vdW}$ (with $E_{\rm b} \sim -P H R^2$) over $H$ and $R$, we find
\begin{equation}
H = c_1 R\bigl ({\gamma}/{Y_0}\bigr )^{1/4}, \qquad
P = c_2 \bigl(\gamma^3 Y_0\bigr)^{1/4}/R . 
\label{eq:standard:appr}
\end{equation}
Here the coefficients $c_1 \approx 0.86$ and $c_2\approx 1.84$ has been obtained from the approximate solution of the Euler-Lagrange equations for $h(r)$ and $\bm{u}(r)$ \cite{Yue2012}.
The results \eqref{eq:standard:appr} are applicable under conditions $\gamma\ll Y_0$ and $R\gg a (Y_0/\gamma)^{1/4}$ which guarantee $H \ll R$ and $H\gg a$, respectively.
We note that the minimization of the energy implies that $|E_{\rm b}| \sim E_{\rm vdW}$ which gives $P\sim \gamma/H$. This relation between $P$ and $H$ will hold for all considered regimes below. Therefore, in what follows we shall present expressions for the maximal height $H$ only.

\textsf{The effect of thermal fluctuations.}\ --- A finite temperature induces the thermal fluctuations of the membrane. These thermal fluctuations are essentially the in-plane and flexural (out-of-plane) phonons. The in-plane phonons  induce the long-ranged interaction between flexural phonons, Eq. \eqref{eq:elastic:energy:mod}. The most dangerous are the out-of-plane phonons with wave vectors $q< 1/R_*$ \cite{Nelson1987,Aronovitz1988}, where $R_* \sim  \varkappa_0/\sqrt{Y_0 T}$ is the so-called Ginzburg length \cite{Aronovitz1989}. Therefore, at finite temperature for the bubble of radius $R>R_*$  one needs to integrate out the flexural phonons with momenta $1/R<q<1/R_*$ before derivation of the Euler-Lagrange equation for $h$. Essentially, integration over the out-of-plane phonons leads to the same form of the bending and elastic energies as given by Eqs. \eqref{eq:bending:energy:mod} and \eqref{eq:elastic:energy:mod} but with the renormalized bending rigidity and Young's modulus \cite{Aronovitz1988}:
\begin{equation}
\varkappa(R) = \varkappa_0\bigl (R/R_*\bigr)^\eta, \qquad Y(R)=Y_0 \bigl(R/R_*\bigr)^{-2+2\eta} .
\label{eq:renorm}
\end{equation}
Here $\eta$ is the universal exponent which depends on the dimensionality of a membrane and of an embedded space. For the clean 2D crystalline membrane in three-dimensional space numerics predicts $\eta\approx 0.8$ \cite{Bowick1996,Troster2013}.

The presence of a non-zero tension $\sigma$ affects the thermal fluctuations. There is the characteristic tension $\sigma_* = \varkappa_0/R_*^2 \sim TY_0/\varkappa_0$
\cite{Roldan2011,Kosmrlj2016,Gornyi2016,Burmistrov2016}.
For $\sigma \ll \sigma_*$ the scaling \eqref{eq:renorm} holds for the interval $R_*\ll R\ll R_\sigma$ where $R_\sigma = R_* (\sigma/\sigma_*)^{1/(2-\eta)}$ is the solution of the equation $\sigma = \varkappa(R_\sigma)/R_\sigma^2$. For $R\gg R_\sigma$ the bending rigidity and the Young's modulus saturates at the values $\varkappa(R_\sigma)$ and $Y(R_\sigma)$, respectively.\!\footnote{Here we neglect weak logarithmic dependence on ${R}$ of the bending rigidity for ${R\gg R_\sigma}$ (cf. {R}ef. \cite{Kosmrlj2016}). {S}uch logarithmic corrections are beyond accuracy of our estimates.}  
For $\sigma > \sigma_*$ ($R_\sigma<R_*$) the thermal fluctuations are completely suppressed and at finite temperature one can minimize the unrenormalized bending, Eq. \eqref{eq:bending:energy:mod},
and elastic, Eq. \eqref{eq:elastic:energy:mod}, energies.

The pressure $P$ inside the bubble results in a non-zero tension $\sigma_P \sim P R_0$ where   
$R_0 \sim R^2/H$ is the radius of the curvature of the membrane on the bubble \cite{Paulose2012}. Using Eq. \eqref{eq:standard:appr}, we find $\sigma_P \sim \sqrt{\gamma Y_0}$. Such tension is enough to suppress 
the thermal fluctuations provided $\sigma_P \gg \sigma_*$, i.e. the standard approach is only valid at low enough temperatures:
$T \ll  T_\gamma \sim \varkappa_0 \sqrt{\gamma/Y_0}$ .
The energy scale $T_\gamma$ has clear physical meaning of  the temperature at which the van der Waals energy for the bubble of radius $R_*$ becomes of the order of the temperature. In other words, for $T \gg T_\gamma$ the van der Waals energy does not suppress the thermal fluctuations. Below we shall study the high temperature regime, $T \gg T_\gamma$.

\textsf{Bending dominated regime.}\ --- We start from the bubble with the radius $R\ll R_*$. At such small lengthscale there is no renormalization of the bending rigidity and Young's modulus. However, as follows from above, the standard approach cannot be correct. The only resolution is the assumption that the bending energy is dominated over elastic one, $E_{\rm bend} \gg E_{\rm el}$, i.e. $H \ll a$. After minimization of $E_{\rm bend}+E_{\rm b}+E_{\rm vdW}$ over $H$ and $R$, we find
\begin{equation}
H = c_3 a \bigl (T_\gamma/T\bigr ) \bigl (R/R_*\bigr)^2 ,
\qquad R\ll R_* .
\label{eq:bending:dominated:short}
\end{equation}
Here $c_3\approx 0.65$ is found from solution of the Euler-Lagrange equation for $h(r)$ (Supplemental Material ~\cite{SM}). 
 
Now we assume that the bubble radius $R\gg R_*$. At such lengthscales one has to take into account the renormalization of the bending rigidity and Young's modulus (provided the scale $R_\sigma$ is large enough). The renormalization changes the estimates for the bending and elastic energies:
$E_{\rm bend} \sim \varkappa(R) {H^2}/{R^2}$ and $E_{\rm el} \sim Y(R) {H^4}/{R^2}$.
Again we assume that the bending energy is larger than the elastic one, $E_{\rm bend}\gg E_{\rm el}$. This implies that $H\ll a (R/R_*)^{1-\eta/2}$. Minimization of $E_{\rm bend}+E_{\rm b}+E_{\rm vdW}$ yields 
\begin{equation}
H  = c_4 a \bigl (T_\gamma/T\bigr ) \bigl(R/R_*\bigr)^{2-\frac{\eta}{2}}, \,\, R_*\ll R\ll R_* T/T_\gamma ,
\label{eq:bending:dominated12}
\end{equation}
where $c_4\approx 0.90$ \cite{SM}. 
The upper bound on $R$ in Eq. \eqref{eq:bending:dominated12} comes from the condition  $E_{\rm bend} \gg E_{\rm el}$. In the above analysis we did not take into account the tension of the membrane due to the pressure. Using Eq.  \eqref{eq:bending:dominated12}, we find
the following estimate: $\sigma_P \sim \gamma (R/H)^2\sim \sigma_* (R_*/R)^{2-\eta}$, i.e. $R_\sigma \sim R$. Since the power-law renormalization \eqref{eq:renorm} is caused by the flexural phonons with momentum $q>1/R$ the tension $\sigma_P$ is indeed irrelevant for the thermal fluctuations in the regime $R_*\ll R \ll R_* T/T_\gamma$.  

\begin{figure}
\includegraphics[width=.85\linewidth]{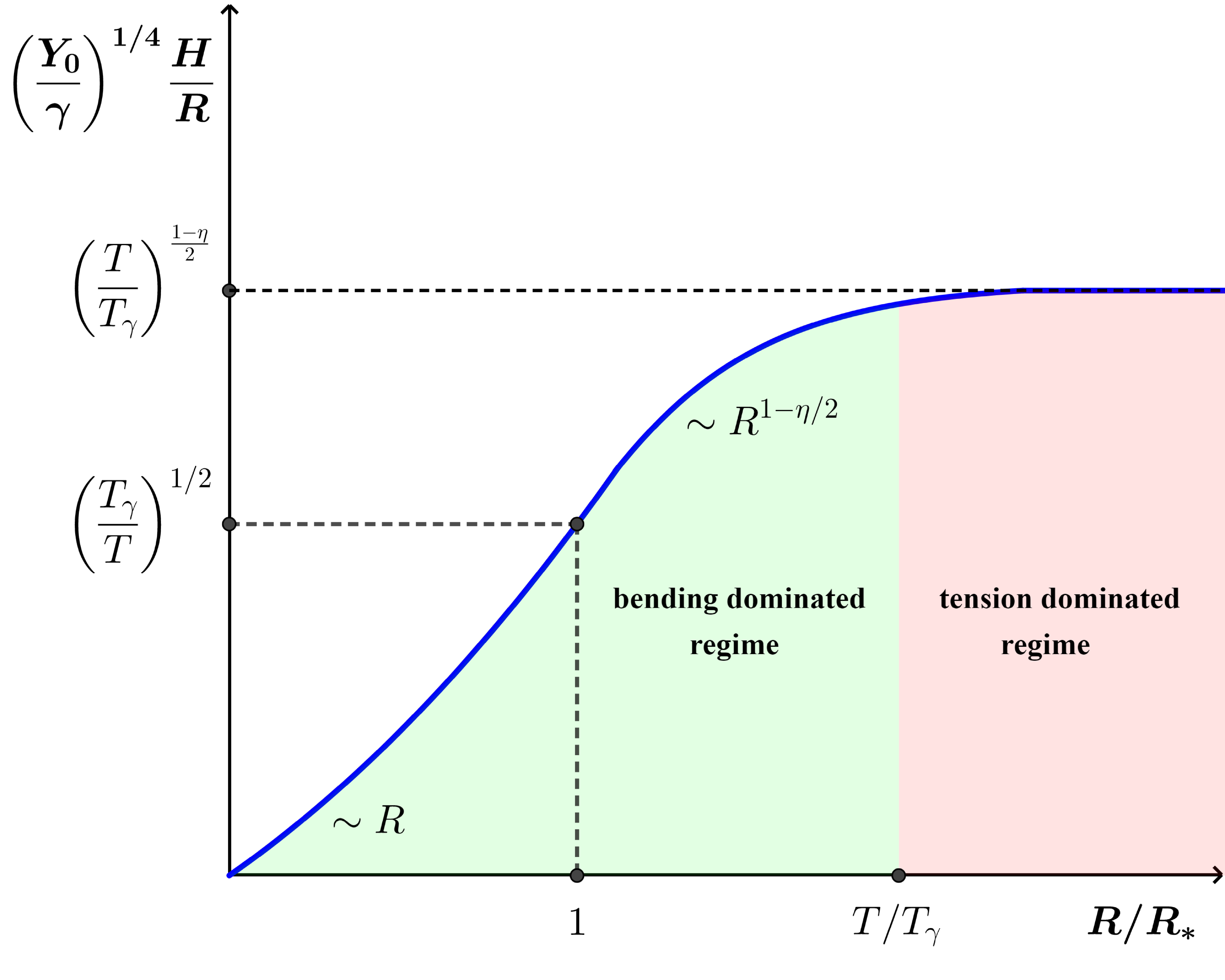}
\caption{The dependence of the aspect ratio on the radius of the bubble at high temperatures, $T\gg T_\gamma$.}
\label{Figure:Height}
\end{figure}

\textsf{Tension dominated regime.}\ ---  For the bubbles of radius $R\gg R_* T/T_\gamma$ the elastic energy is dominated over the bending one, 
$E_{\rm el} \gg E_{\rm bend}$. Then the minimization of $E_{\rm el}+E_{\rm b}+E_{\rm vdW}$ over $H$ and $R$ implies that $E_{\rm el}\sim |E_{\rm b}| \sim E_{\rm vdW}$. Therefore, the pressure-induced tension $\sigma_P \sim |E_{\rm b}|/H^2\sim E_{\rm el}/H^2 \gg E_{\rm bend}/H^2$. This estimate means that the pressure induced tension is important and the corresponding length scale is short, $R_\sigma \ll R$. In such regime the bending rigidity and Young's modulus are independent of $R$ albeit strongly renormalized. Therefore, we can use the results of the standard approach but with the Young's modulus $Y(R_\sigma)$ instead of $Y_0$. In particular, the tension induced by the pressure is given as $\sigma_P \sim \sqrt{\gamma Y(R_\sigma)}$. Hence the length scale $R_\sigma$ satisfies the following equation: $\varkappa(R_\sigma)/R_\sigma^2=\sqrt{\gamma Y(R_\sigma)}$. Its solution yields
$R_\sigma \sim R_* T/T_\gamma$. This justifies that the profile of the bubbles with $R\gg R_* T/T_\gamma$ are governed by pressure induced tension. The characteristic radius $R_\sigma$ has simple physical meaning. The bubble of such radius has the adhesion energy, $\pi \gamma R_\sigma^2$, equal to $T$. Using Eq. \eqref{eq:standard:appr} with the renormalized Young's modulus, we find 
\begin{equation}
H = c_1 a \bigl (R/R_*\bigr ) \bigl(T/T_\gamma\bigr)^{-\eta/2}, \quad R_*T/T_\gamma \ll R .
\label{eq:elastic:dominated1} 
\end{equation}
We mention that although the aspect ratio $H/R$ of the bubbles with $R\gg R_*T/T_\gamma$ is independent of $R$, it is not the constant but depends on the temperature. 
The value of the aspect ratio is much larger than one would predict on the basis of the standard approach. The behavior of the aspect ratio on $R$ at $T\gg T_\gamma$ is shown in Fig. \ref{Figure:Height}. 
 
\textsf{The anomalous thermodynamics.}\ --- The temperature dependence of the maximal height $H$ depends on the equation of state of the substance inside the bubble. 
We start from the case of the liquid  bubble. Then we can approximate the equation of state 
by the constant volume condition: $V={\rm const}$. We assume that the bubble has large enough volume, $V\gg V_\gamma \sim a^3 \varkappa_0/T_\gamma$. In the opposite case, $V\ll V_\gamma$,  
$H$ is smaller than the effective thickness of the membrane at all temperatures \cite{SM}.

\begin{figure}
\centering
\includegraphics[width=.85\linewidth]{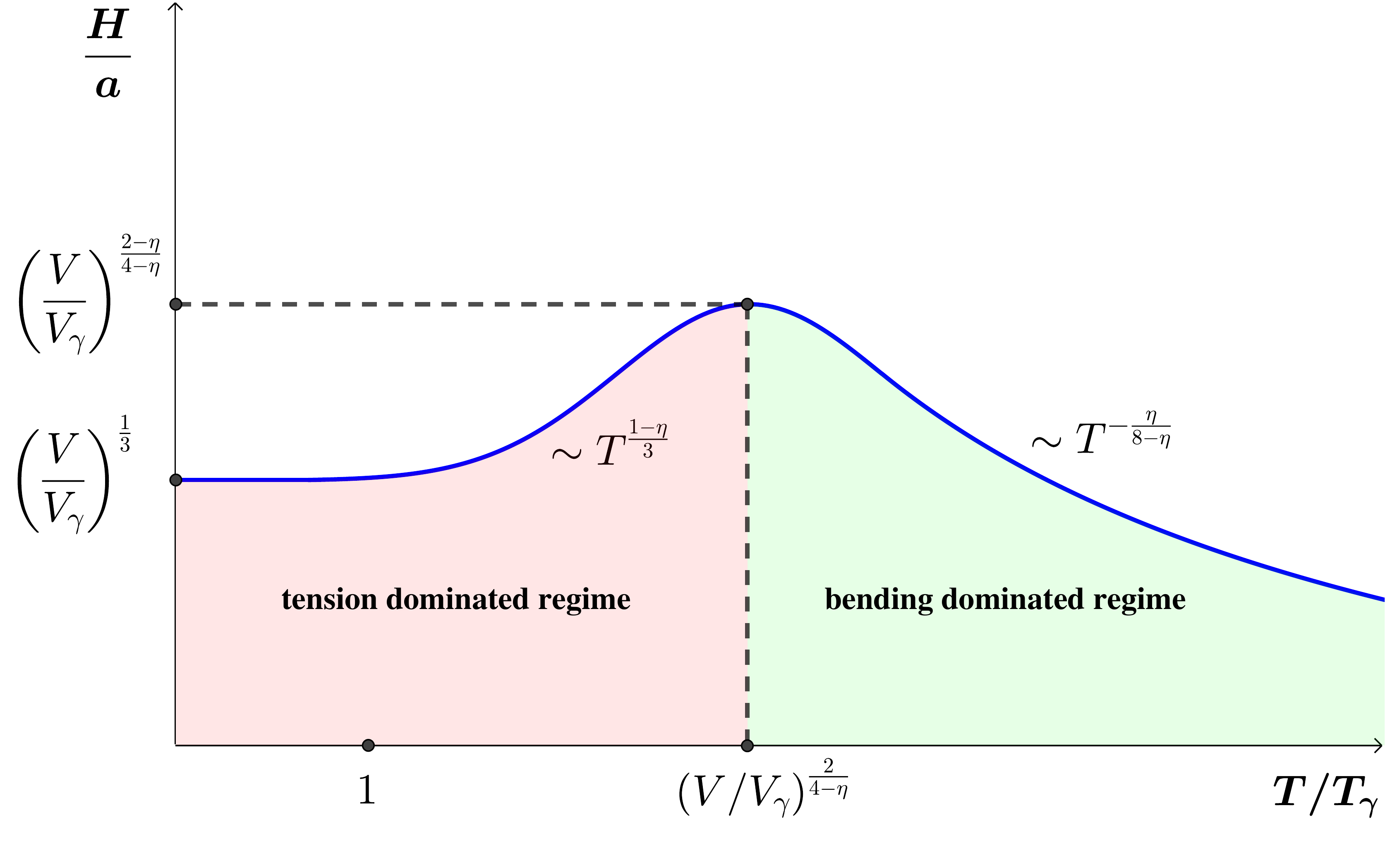}
\caption{The dependence of the maximal height of the bubble with liquid on the temperature for the case of large volume, $V\gg V_\gamma$.}
\label{Figure:Liquid}
\end{figure}

At low temperatures, $T\ll T_\gamma$, the maximal height is given by Eq. \eqref{eq:standard:appr}, i.e. $H$ is independent of $T$: 
\begin{equation}
H \sim a \bigl (V/V_\gamma\bigr )^{1/3},\qquad T
\ll T_\gamma  .
\label{eq:H:reg21}
\end{equation} 
At 
$T_\gamma\ll T\ll T_\gamma (V / V_\gamma)^{\frac{2}{4-\eta}}$ the thermal fluctuations are important but the physics is dominated by the tension induced by the pressure. Using Eq. \eqref{eq:elastic:dominated1}, we find 
\begin{equation}
H \sim a \left (\frac{V T^{1-\eta}}{V_\gamma T_\gamma^{1-\eta}}\right )^{\frac{1}{3}}
,\, 
T_\gamma \ll
T \ll T_\gamma \left (\frac{V}{V_\gamma}\right )^{\frac{2}{4-\eta}} . 
\label{eq:H:reg23}
\end{equation} 
At high temperatures, $T\gg T_\gamma (V / V_\gamma)^{\frac{2}{4-\eta}}$, the maximal height of the bubble is described by the theory of the bending dominated regime, Eq. \eqref{eq:bending:dominated12}. Then, we find that $H$ is decreasing with increase of temperature:
\begin{equation}
H \sim a \left (\frac{V^{4-\eta} T^\eta_\gamma}{V_\gamma^{4-\eta}T^\eta}\right )^\frac{1}{8-\eta}
,\qquad 
T_\gamma \left (\frac{V}{V_\gamma}\right )^{\frac{2}{4-\eta}}\ll T  .
\label{eq:H:reg23}
\end{equation} 
Therefore, in the regime of large volumes, $V\gg V_\gamma$, the maximal height of the bubble has non-monotonous dependence on temperature with the maximum 
at temperature $T_{\rm max}\sim T_\gamma (V/V_\gamma)^{2/(4-\eta)}$ (see Fig. \ref{Figure:Liquid}). The non-monotonous dependence of $H$ implies the change of the sign of the linear thermal expansion coefficient $\alpha_H$ at temperature $T_{\rm max}$:
\begin{equation}
\alpha_H = \frac{1}{T}\begin{cases}
0, & T\ll T_\gamma , \\
\frac{1-\eta}{3}, & T_\gamma\ll T\ll T_{\rm max}, \\
-\frac{\eta}{8-\eta}, & T_{\rm max}\ll T .
\end{cases}
\label{eq:LTEC}
\end{equation}
Therefore, by measuring the slope of $\alpha_H$ against $1/T$ one can extract the bending rigidity exponent of the membrane. The result \eqref{eq:LTEC} is derived with the neglect of temperature dependence of the adhesion energy.

Now we discuss the case of the bubble with a gas inside. For a sake of simplicity, we use the equation of state of the ideal gas, $PV=N T$, where $N$ is the number of atoms of the gas. Using the relations $V\sim HR^2\sim \gamma R^2/P$, we find that the radius of the bubble with the ideal gas is always given as $R\sim \sqrt{N T/\gamma}$. At low temperature $T\ll T_\gamma$, using  Eq. \eqref{eq:standard:appr}, we find that the maximal height of the bubble grows with temperature as
\begin{equation}
H \sim a \sqrt{N} \bigl (T/T_\gamma\bigr )^{1/2}, \qquad T\ll T_\gamma .
\label{eq:H:gas:1}
\end{equation}  
We note that, strictly speaking, this estimate is valid for $T\gg T_\gamma/N$. Under assumption of the macroscopic number of atoms, $N\gg 1$, inside the bubble this limitation on $T$ is completely irrelevant.   

\begin{figure}
\centering
\includegraphics[width=.85\linewidth]{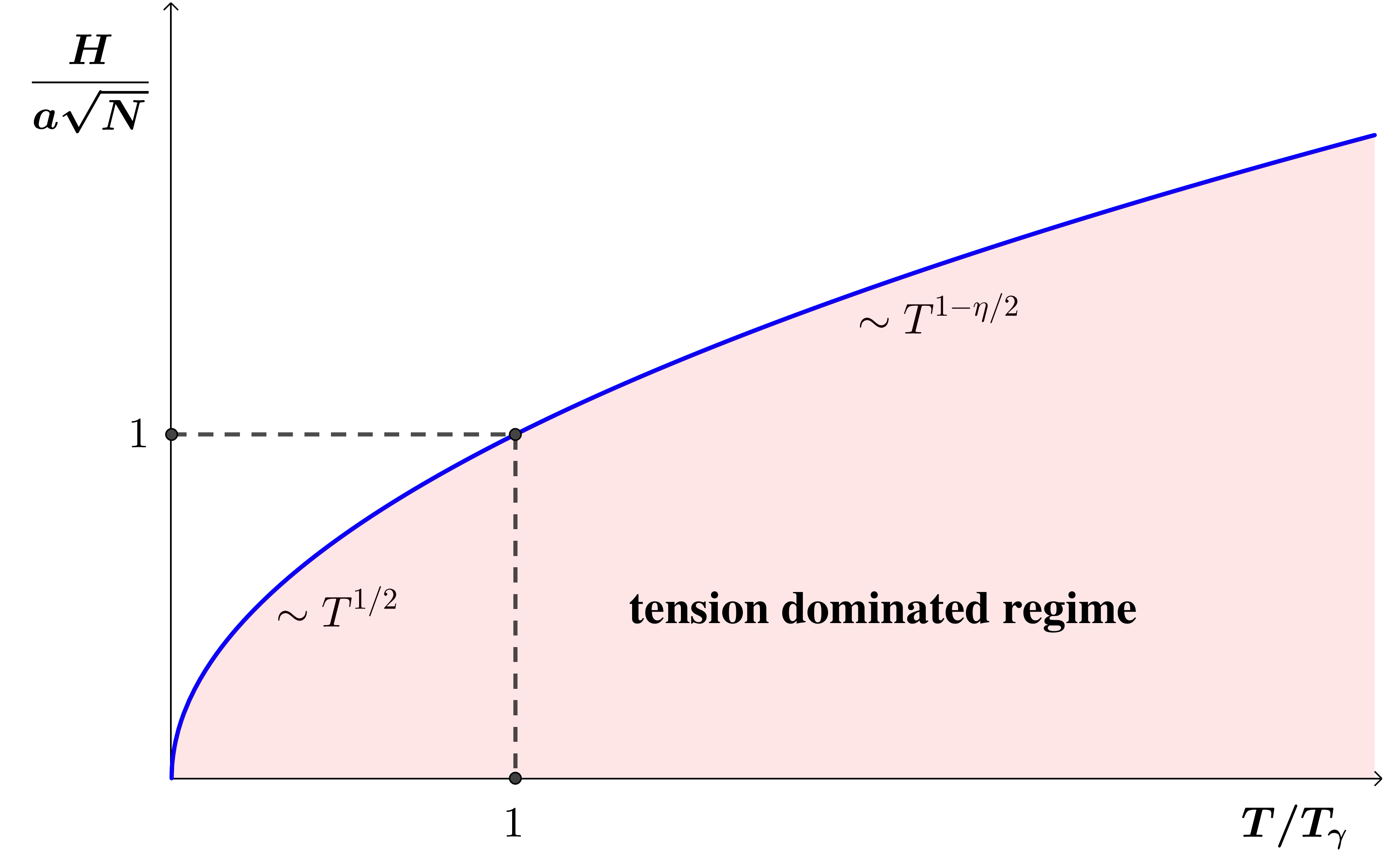}
\caption{The dependence of the maximal height of the bubble with the ideal gas inside on the temperature.}
\label{Figure:Gas}
\end{figure}

For high temperatures, $T\gg T_\gamma$, the bubbles of radius $R\gg R_* T/T_\gamma$ (this condition is equivalent to the condition $N\gg 1$) can be formed only. Therefore, the bubble is in the tension dominated regime such that its maximal height is described by Eq. \eqref{eq:elastic:dominated1}. Then, we obtain
\begin{equation}
H \sim a \sqrt{N} \bigl (T/T_\gamma\bigr )^{1-\eta/2}, \qquad T\gg T_\gamma .
\label{eq:H:gas:2}
\end{equation}
The above results show that the maximal height of the bubble with the ideal gas inside is the monotonously growing function of temperature (see Fig. \ref{Figure:Gas}). Therefore, the linear thermal expansion coefficient is always positive:
\begin{equation}
\alpha_H = \frac{1}{T} \begin{cases}
1/2, & \qquad T\ll T_\gamma , \\
1-\eta/2, & \qquad T_\gamma\ll T .
\end{cases}
\end{equation}
As in the case of the bubble with a liquid the slope of $\alpha_H$ with $1/T$ allows to extract the value of the exponent $\eta$.

\textsf{Discussion and summary.}\ --- With the use of a more realistic equation of state one can compute the temperature dependence of the maximal height along the liquid-to-gas isotherm. In particular, one can analyze how the anomalous elasticity affects the liquid-to-gas transition in the bubble. This phenomenon has been studied recently in Ref. \cite{Zhilyaev2019} but within the standard approach which ignores the thermal fluctuations of the membrane.  

It is known \cite{Paulose2012,Kosmrlj2017} that the anomalous elasticity of membrane affects the stability of spherical membrane shells. The renormalization of the bending rigidity and Young's modulus decreases the pressure induced tension towards a negative value which is enough for developing the buckling instability. In principle, similar mechanism of instability is also applicable for the curved membrane above the bubble. However, our estimates indicate that such buckling instability is out of reach \cite{SM}.  

Also it is worthwhile to mention that in the bending dominated regime there are large thermodynamics fluctuations of the bubble height which might complicate the experimental observation of the predicted dependence of the average height of the bubble on $R$ and $T$ \cite{SM}.

Let us estimate the relevant parameters for our theory in the case of a van der Waals heterostructure made of a graphene monolayer on a monolayer of hBN. Using the known values of Young's modulus, bending rigidity, and the effective thickness of the graphene: $Y_0 \approx 22$ eV \AA$^{-2}$,  
$\varkappa_0 \approx 1.1$ eV, $a\approx 0.6$ \AA, we can estimate the Ginzburg length as $R_* \approx  4$~\AA \, at $T=300$ K \cite{Roldan2011}. Assuming that 
the total adhesive energy is dominated by the adhesive energy between graphene and hBN, $\gamma \approx \gamma_{\rm ms}\approx 0.008$ eV \AA$^{-2}$ \cite{Megra2019}, we find  $T_\gamma \approx 250$ K. This estimate indicates that the aspect ratio of the bubbles between graphene and hBN measured recently \cite{Khestanova2016}  can be described by our theory in the tension dominated regime, Eq. \eqref{eq:elastic:dominated1}. Using the experimentally observed aspect ratio $0.11$ we obtain $T_\gamma \approx 220$ K and the adhesion energy $\gamma \approx 0.007$ eV \AA$^{-2}$. The later is 20 per cent higher than the value extracted in Ref. \cite{Khestanova2016} within the standard approach, Eq. \eqref{eq:standard:appr}. This implies that the proper account for the thermal fluctuations can be crucial for the precision measurements of the adhesion energy via the pressurised blister test. The characteristic volume for the bubble between graphene and hBN can be estimated as  $V_\gamma \approx 10$ \AA$^3$. Such smallness of the value of $V_\gamma$ suggests that the non-monotonous behavior of the maximal height on temperature in graphene-on-hBN structure could be observed experimentally for the liquid bubbles of few nanometer radius only. 

In summary, we have demonstrated that the anomalous elasticity of membranes affects the profile of a bubble in van der Waals heterostructures at high temperatures.
We have shown that the renormalization of the bending rigidity and Young's modulus results in the anomalous dependence of the maximal height of the bubble on its radius and temperature. 
Our estimates suggest that for graphene-based van der Waals heterostructures the anomalous regime predicted in the paper is experimentally accessible at ambient conditions. 

\textsf{Acknowledgements.}\ ---{A.A.L. and S.S.B. are grateful to the ``FIZIKA'' foundation for support during the 2019 Landau Institute summer school on theoretical physics where this project was initiated. I.S.B. is grateful to I. Gornyi, V. Kachorovskii, and A. Mirlin for very useful discussions as well as to I. Kolokolov and K. Tikhonov for valuable remarks. The work was partially supported by the programs of the Russian Ministry of Science and Higher Education and by the Basic Research Program of HSE.}

\bibliography{biblio-graphene}

\begin{thebibliography}{39}%
\makeatletter
\providecommand \@ifxundefined [1]{%
 \@ifx{#1\undefined}
}%
\providecommand \@ifnum [1]{%
 \ifnum #1\expandafter \@firstoftwo
 \else \expandafter \@secondoftwo
 \fi
}%
\providecommand \@ifx [1]{%
 \ifx #1\expandafter \@firstoftwo
 \else \expandafter \@secondoftwo
 \fi
}%
\providecommand \natexlab [1]{#1}%
\providecommand \enquote  [1]{``#1''}%
\providecommand \bibnamefont  [1]{#1}%
\providecommand \bibfnamefont [1]{#1}%
\providecommand \citenamefont [1]{#1}%
\providecommand \href@noop [0]{\@secondoftwo}%
\providecommand \href [0]{\begingroup \@sanitize@url \@href}%
\providecommand \@href[1]{\@@startlink{#1}\@@href}%
\providecommand \@@href[1]{\endgroup#1\@@endlink}%
\providecommand \@sanitize@url [0]{\catcode `\\12\catcode `\$12\catcode
  `\&12\catcode `\#12\catcode `\^12\catcode `\_12\catcode `\%12\relax}%
\providecommand \@@startlink[1]{}%
\providecommand \@@endlink[0]{}%
\providecommand \url  [0]{\begingroup\@sanitize@url \@url }%
\providecommand \@url [1]{\endgroup\@href {#1}{\urlprefix }}%
\providecommand \urlprefix  [0]{URL }%
\providecommand \Eprint [0]{\href }%
\providecommand \doibase [0]{http://dx.doi.org/}%
\providecommand \selectlanguage [0]{\@gobble}%
\providecommand \bibinfo  [0]{\@secondoftwo}%
\providecommand \bibfield  [0]{\@secondoftwo}%
\providecommand \translation [1]{[#1]}%
\providecommand \BibitemOpen [0]{}%
\providecommand \bibitemStop [0]{}%
\providecommand \bibitemNoStop [0]{.\EOS\space}%
\providecommand \EOS [0]{\spacefactor3000\relax}%
\providecommand \BibitemShut  [1]{\csname bibitem#1\endcsname}%
\let\auto@bib@innerbib\@empty
\bibitem [{\citenamefont {Dai}\ \emph {et~al.}(2019)\citenamefont {Dai},
  \citenamefont {Liu},\ and\ \citenamefont {Zhang}}]{Dai2019}%
  \BibitemOpen
  \bibfield  {author} {\bibinfo {author} {\bibfnamefont {Z.}~\bibnamefont
  {Dai}}, \bibinfo {author} {\bibfnamefont {L.}~\bibnamefont {Liu}}, \ and\
  \bibinfo {author} {\bibfnamefont {Z.}~\bibnamefont {Zhang}},\ }\bibfield
  {title} {\enquote {\bibinfo {title} {Strain engineering of 2d materials:
  Issues and opportunities at the interface},}\ }\href
  {https://doi.org/10.1002/adma.201805417} {\bibfield  {journal} {\bibinfo
  {journal} {Adv. Mater.}\ ,\ \bibinfo {pages} {1805417}} (\bibinfo {year}
  {2019})}\BibitemShut {NoStop}%
\bibitem [{\citenamefont {Megra}\ and\ \citenamefont {Suk}(2019)}]{Megra2019}%
  \BibitemOpen
  \bibfield  {author} {\bibinfo {author} {\bibfnamefont {Y.~T.}\ \bibnamefont
  {Megra}}\ and\ \bibinfo {author} {\bibfnamefont {J.~W.}\ \bibnamefont
  {Suk}},\ }\bibfield  {title} {\enquote {\bibinfo {title} {Adhesion properties
  of 2d materials},}\ }\href {\doibase 10.1088/1361-6463/ab27ad} {\bibfield
  {journal} {\bibinfo  {journal} {J. Phys. D: Appl. Phys.}\ }\textbf {\bibinfo
  {volume} {52}},\ \bibinfo {pages} {364002} (\bibinfo {year}
  {2019})}\BibitemShut {NoStop}%
\bibitem [{\citenamefont {Haigh}\ \emph {et~al.}(2012)\citenamefont {Haigh},
  \citenamefont {Gholinia}, \citenamefont {Jalil}, \citenamefont {Romani},
  \citenamefont {Britnell}, \citenamefont {Elias}, \citenamefont {Novoselov},
  \citenamefont {Ponomarenko}, \citenamefont {Geim},\ and\ \citenamefont
  {Gorbachev}}]{Haigh2012}%
  \BibitemOpen
  \bibfield  {author} {\bibinfo {author} {\bibfnamefont {S.~J.}\ \bibnamefont
  {Haigh}}, \bibinfo {author} {\bibfnamefont {A.}~\bibnamefont {Gholinia}},
  \bibinfo {author} {\bibfnamefont {R.}~\bibnamefont {Jalil}}, \bibinfo
  {author} {\bibfnamefont {S.}~\bibnamefont {Romani}}, \bibinfo {author}
  {\bibfnamefont {L.}~\bibnamefont {Britnell}}, \bibinfo {author}
  {\bibfnamefont {D.~C.}\ \bibnamefont {Elias}}, \bibinfo {author}
  {\bibfnamefont {K.~S.}\ \bibnamefont {Novoselov}}, \bibinfo {author}
  {\bibfnamefont {L.~A.}\ \bibnamefont {Ponomarenko}}, \bibinfo {author}
  {\bibfnamefont {A.~K.}\ \bibnamefont {Geim}}, \ and\ \bibinfo {author}
  {\bibfnamefont {R.}~\bibnamefont {Gorbachev}},\ }\bibfield  {title} {\enquote
  {\bibinfo {title} {Cross-sectional imaging of individual layers and buried
  interfaces of graphene-based heterostructures and superlattices},}\ }\href
  {\doibase 10.1038/nmat3386} {\bibfield  {journal} {\bibinfo  {journal} {Nat.
  Mater.}\ }\textbf {\bibinfo {volume} {11}},\ \bibinfo {pages} {764} (\bibinfo
  {year} {2012})}\BibitemShut {NoStop}%
\bibitem [{\citenamefont {Khestanova}\ \emph {et~al.}(2016)\citenamefont
  {Khestanova}, \citenamefont {Guinea}, \citenamefont {Fumagalli},
  \citenamefont {Geim},\ and\ \citenamefont {Grigorieva}}]{Khestanova2016}%
  \BibitemOpen
  \bibfield  {author} {\bibinfo {author} {\bibfnamefont {E.}~\bibnamefont
  {Khestanova}}, \bibinfo {author} {\bibfnamefont {F.}~\bibnamefont {Guinea}},
  \bibinfo {author} {\bibfnamefont {L.}~\bibnamefont {Fumagalli}}, \bibinfo
  {author} {\bibfnamefont {A.~K.}\ \bibnamefont {Geim}}, \ and\ \bibinfo
  {author} {\bibfnamefont {I.~V.}\ \bibnamefont {Grigorieva}},\ }\bibfield
  {title} {\enquote {\bibinfo {title} {Universal shape and pressure inside
  bubbles appearing in van der waals heterostructures},}\ }\href
  {http://dx.doi.org/10.1038/ncomms12587} {\bibfield  {journal} {\bibinfo
  {journal} {Nat. Commun.}\ }\textbf {\bibinfo {volume} {7}},\ \bibinfo {pages}
  {12587} (\bibinfo {year} {2016})}\BibitemShut {NoStop}%
\bibitem [{\citenamefont {Dai}\ \emph {et~al.}(2018)\citenamefont {Dai},
  \citenamefont {Hou}, \citenamefont {Sanchez}, \citenamefont {Wang},
  \citenamefont {Brennan}, \citenamefont {Zhang}, \citenamefont {Liu},\ and\
  \citenamefont {Lu}}]{Dai2018}%
  \BibitemOpen
  \bibfield  {author} {\bibinfo {author} {\bibfnamefont {Z.}~\bibnamefont
  {Dai}}, \bibinfo {author} {\bibfnamefont {Y.}~\bibnamefont {Hou}}, \bibinfo
  {author} {\bibfnamefont {D.~A.}\ \bibnamefont {Sanchez}}, \bibinfo {author}
  {\bibfnamefont {G.}~\bibnamefont {Wang}}, \bibinfo {author} {\bibfnamefont
  {C.~J.}\ \bibnamefont {Brennan}}, \bibinfo {author} {\bibfnamefont
  {Z.}~\bibnamefont {Zhang}}, \bibinfo {author} {\bibfnamefont
  {L.}~\bibnamefont {Liu}}, \ and\ \bibinfo {author} {\bibfnamefont
  {N.}~\bibnamefont {Lu}},\ }\bibfield  {title} {\enquote {\bibinfo {title}
  {Interface-governed deformation of nanobubbles and nanotents formed by
  two-dimensional materials},}\ }\href {\doibase
  10.1103/PhysRevLett.121.266101} {\bibfield  {journal} {\bibinfo  {journal}
  {Phys. Rev. Lett.}\ }\textbf {\bibinfo {volume} {121}},\ \bibinfo {pages}
  {266101} (\bibinfo {year} {2018})}\BibitemShut {NoStop}%
\bibitem [{\citenamefont {Ghodsi}\ \emph {et~al.}(2019)\citenamefont {Ghodsi},
  \citenamefont {Megaridis}, \citenamefont {Shahbazian-Yassar},\ and\
  \citenamefont {Shokuhfar}}]{Ghodsi2019}%
  \BibitemOpen
  \bibfield  {author} {\bibinfo {author} {\bibfnamefont {S.M.}\ \bibnamefont
  {Ghodsi}}, \bibinfo {author} {\bibfnamefont {C.~M.}\ \bibnamefont
  {Megaridis}}, \bibinfo {author} {\bibfnamefont {R.}~\bibnamefont
  {Shahbazian-Yassar}}, \ and\ \bibinfo {author} {\bibfnamefont
  {T.}~\bibnamefont {Shokuhfar}},\ }\bibfield  {title} {\enquote {\bibinfo
  {title} {Advances in graphene-based liquid cell electron microscopy: Working
  principles, opportunities, and challenges},}\ }\href
  {https://doi.org/10.1002/smtd.201900026} {\bibfield  {journal} {\bibinfo
  {journal} {Small Methods}\ ,\ \bibinfo {pages} {1900026}} (\bibinfo {year}
  {2019})}\BibitemShut {NoStop}%
\bibitem [{\citenamefont {Tyurnina}\ \emph {et~al.}(2019)\citenamefont
  {Tyurnina}, \citenamefont {Bandurin}, \citenamefont {Khestanova},
  \citenamefont {Koperski}, \citenamefont {Guinea}, \citenamefont {Grigorenko},
  \citenamefont {Geim},\ and\ \citenamefont {Grigorieva}}]{Tyurnina2019}%
  \BibitemOpen
  \bibfield  {author} {\bibinfo {author} {\bibfnamefont {A.~V.}\ \bibnamefont
  {Tyurnina}}, \bibinfo {author} {\bibfnamefont {D.~A.}\ \bibnamefont
  {Bandurin}}, \bibinfo {author} {\bibfnamefont {V.~G.}\ \bibnamefont
  {Khestanova}, \bibfnamefont {E.and~Kravets}}, \bibinfo {author}
  {\bibfnamefont {M.}~\bibnamefont {Koperski}}, \bibinfo {author}
  {\bibfnamefont {F.}~\bibnamefont {Guinea}}, \bibinfo {author} {\bibfnamefont
  {A.~N.}\ \bibnamefont {Grigorenko}}, \bibinfo {author} {\bibfnamefont
  {A.~K.}\ \bibnamefont {Geim}}, \ and\ \bibinfo {author} {\bibfnamefont
  {I.~V.}\ \bibnamefont {Grigorieva}},\ }\bibfield  {title} {\enquote {\bibinfo
  {title} {Strained bubbles in van der waals heterostructures as local emitters
  of photoluminescence with adjustable wavelength},}\ }\href {\doibase
  10.1021/acsphotonics.8b01497} {\bibfield  {journal} {\bibinfo  {journal} {ACS
  Photonics}\ }\textbf {\bibinfo {volume} {6}},\ \bibinfo {pages} {516}
  (\bibinfo {year} {2019})}\BibitemShut {NoStop}%
\bibitem [{\citenamefont {Koenig}\ \emph {et~al.}(2011)\citenamefont {Koenig},
  \citenamefont {Boddeti}, \citenamefont {Dunn},\ and\ \citenamefont
  {Bunch}}]{Bunch2011}%
  \BibitemOpen
  \bibfield  {author} {\bibinfo {author} {\bibfnamefont {S.~P.}\ \bibnamefont
  {Koenig}}, \bibinfo {author} {\bibfnamefont {N.~G.}\ \bibnamefont {Boddeti}},
  \bibinfo {author} {\bibfnamefont {M.~L.}\ \bibnamefont {Dunn}}, \ and\
  \bibinfo {author} {\bibfnamefont {J.~S.}\ \bibnamefont {Bunch}},\ }\bibfield
  {title} {\enquote {\bibinfo {title} {Ultrastrong adhesion of graphene
  membranes},}\ }\href {https://www.nature.com/articles/nnano.2011.123}
  {\bibfield  {journal} {\bibinfo  {journal} {Nat. Nanotechnol.}\ }\textbf
  {\bibinfo {volume} {6}},\ \bibinfo {pages} {543} (\bibinfo {year}
  {2011})}\BibitemShut {NoStop}%
\bibitem [{\citenamefont {Boddeti}\ \emph {et~al.}(2013)\citenamefont
  {Boddeti}, \citenamefont {Liu}, \citenamefont {Long}, \citenamefont {Xiao},
  \citenamefont {Bunch},\ and\ \citenamefont {Dunn}}]{Boddeti2013}%
  \BibitemOpen
  \bibfield  {author} {\bibinfo {author} {\bibfnamefont {Narasimha~G.}\
  \bibnamefont {Boddeti}}, \bibinfo {author} {\bibfnamefont {Xinghui}\
  \bibnamefont {Liu}}, \bibinfo {author} {\bibfnamefont {Rong}\ \bibnamefont
  {Long}}, \bibinfo {author} {\bibfnamefont {Jianliang}\ \bibnamefont {Xiao}},
  \bibinfo {author} {\bibfnamefont {J.~Scott}\ \bibnamefont {Bunch}}, \ and\
  \bibinfo {author} {\bibfnamefont {Martin~L.}\ \bibnamefont {Dunn}},\
  }\bibfield  {title} {\enquote {\bibinfo {title} {Graphene blisters with
  switchable shapes controlled by pressure and adhesion},}\ }\href {\doibase
  10.1021/nl4036324} {\bibfield  {journal} {\bibinfo  {journal} {Nano Letters}\
  }\textbf {\bibinfo {volume} {13}},\ \bibinfo {pages} {6216--6221} (\bibinfo
  {year} {2013})}\BibitemShut {NoStop}%
\bibitem [{\citenamefont {Lloyd}\ \emph {et~al.}(2017)\citenamefont {Lloyd},
  \citenamefont {Liu}, \citenamefont {Boddeti}, \citenamefont {Cantley},
  \citenamefont {Long}, \citenamefont {Dunn},\ and\ \citenamefont
  {Bunch}}]{Lloyd2017}%
  \BibitemOpen
  \bibfield  {author} {\bibinfo {author} {\bibfnamefont {D.}~\bibnamefont
  {Lloyd}}, \bibinfo {author} {\bibfnamefont {X.}~\bibnamefont {Liu}}, \bibinfo
  {author} {\bibfnamefont {N.}~\bibnamefont {Boddeti}}, \bibinfo {author}
  {\bibfnamefont {L.}~\bibnamefont {Cantley}}, \bibinfo {author} {\bibfnamefont
  {R.}~\bibnamefont {Long}}, \bibinfo {author} {\bibfnamefont {M.~L.}\
  \bibnamefont {Dunn}}, \ and\ \bibinfo {author} {\bibfnamefont {J.~S.}\
  \bibnamefont {Bunch}},\ }\bibfield  {title} {\enquote {\bibinfo {title}
  {Adhesion, stiffness, and instability in atomically thin {MoS$_2$}
  bubbles},}\ }\href {\doibase 10.1021/acs.nanolett.7b01735} {\bibfield
  {journal} {\bibinfo  {journal} {Nano Letters}\ }\textbf {\bibinfo {volume}
  {17}},\ \bibinfo {pages} {5329} (\bibinfo {year} {2017})}\BibitemShut
  {NoStop}%
\bibitem [{\citenamefont {Wang}\ \emph {et~al.}(2019)\citenamefont {Wang},
  \citenamefont {Dai}, \citenamefont {Xiao}, \citenamefont {Feng},
  \citenamefont {Weng}, \citenamefont {Liu}, \citenamefont {Xu}, \citenamefont
  {Huang},\ and\ \citenamefont {Zhang}}]{Wang2019}%
  \BibitemOpen
  \bibfield  {author} {\bibinfo {author} {\bibfnamefont {G.}~\bibnamefont
  {Wang}}, \bibinfo {author} {\bibfnamefont {Z.}~\bibnamefont {Dai}}, \bibinfo
  {author} {\bibfnamefont {J.}~\bibnamefont {Xiao}}, \bibinfo {author}
  {\bibfnamefont {S.}~\bibnamefont {Feng}}, \bibinfo {author} {\bibfnamefont
  {C.}~\bibnamefont {Weng}}, \bibinfo {author} {\bibfnamefont {L.}~\bibnamefont
  {Liu}}, \bibinfo {author} {\bibfnamefont {Z.}~\bibnamefont {Xu}}, \bibinfo
  {author} {\bibfnamefont {R.}~\bibnamefont {Huang}}, \ and\ \bibinfo {author}
  {\bibfnamefont {Z.}~\bibnamefont {Zhang}},\ }\bibfield  {title} {\enquote
  {\bibinfo {title} {Bending of multilayer van der waals materials},}\ }\href
  {\doibase 10.1103/PhysRevLett.123.116101} {\bibfield  {journal} {\bibinfo
  {journal} {Phys. Rev. Lett.}\ }\textbf {\bibinfo {volume} {123}},\ \bibinfo
  {pages} {116101} (\bibinfo {year} {2019})}\BibitemShut {NoStop}%
\bibitem [{\citenamefont {Yue}\ \emph {et~al.}(2012)\citenamefont {Yue},
  \citenamefont {Gao}, \citenamefont {Huang},\ and\ \citenamefont
  {Liechti}}]{Yue2012}%
  \BibitemOpen
  \bibfield  {author} {\bibinfo {author} {\bibfnamefont {K.}~\bibnamefont
  {Yue}}, \bibinfo {author} {\bibfnamefont {W.}~\bibnamefont {Gao}}, \bibinfo
  {author} {\bibfnamefont {R.}~\bibnamefont {Huang}}, \ and\ \bibinfo {author}
  {\bibfnamefont {K.~M.}\ \bibnamefont {Liechti}},\ }\bibfield  {title}
  {\enquote {\bibinfo {title} {Analytical methods for the mechanics of graphene
  bubbles},}\ }\href {http://dx.doi.org/10.1063/1.4759146} {\bibfield
  {journal} {\bibinfo  {journal} {J. Appl. Phys.}\ }\textbf {\bibinfo {volume}
  {112}},\ \bibinfo {pages} {083512} (\bibinfo {year} {2012})}\BibitemShut
  {NoStop}%
\bibitem [{\citenamefont {Wang}\ \emph {et~al.}(2013)\citenamefont {Wang},
  \citenamefont {Gao}, \citenamefont {Cao}, \citenamefont {Liechti},\ and\
  \citenamefont {Huang}}]{Wang2013}%
  \BibitemOpen
  \bibfield  {author} {\bibinfo {author} {\bibfnamefont {P.}~\bibnamefont
  {Wang}}, \bibinfo {author} {\bibfnamefont {W.}~\bibnamefont {Gao}}, \bibinfo
  {author} {\bibfnamefont {Z.}~\bibnamefont {Cao}}, \bibinfo {author}
  {\bibfnamefont {K.~M.}\ \bibnamefont {Liechti}}, \ and\ \bibinfo {author}
  {\bibfnamefont {R.}~\bibnamefont {Huang}},\ }\bibfield  {title} {\enquote
  {\bibinfo {title} {Numerical analysis of circular graphene bubbles},}\ }\href
  {\doibase 10.1115/1.4024169} {\bibfield  {journal} {\bibinfo  {journal} {J.
  Appl. Mechanics}\ }\textbf {\bibinfo {volume} {80}},\ \bibinfo {pages}
  {040905} (\bibinfo {year} {2013})}\BibitemShut {NoStop}%
\bibitem [{\citenamefont {Nelson}\ and\ \citenamefont
  {Peliti}(1987)}]{Nelson1987}%
  \BibitemOpen
  \bibfield  {author} {\bibinfo {author} {\bibfnamefont {D.R.}\ \bibnamefont
  {Nelson}}\ and\ \bibinfo {author} {\bibfnamefont {L.}~\bibnamefont
  {Peliti}},\ }\bibfield  {title} {\enquote {\bibinfo {title} {Fluctuations in
  membranes with crystalline and hexatic order},}\ }\href {\doibase
  10.1051/jphys:019870048070108500} {\bibfield  {journal} {\bibinfo  {journal}
  {Journal de Physique}\ }\textbf {\bibinfo {volume} {48}},\ \bibinfo {pages}
  {1085--1092} (\bibinfo {year} {1987})}\BibitemShut {NoStop}%
\bibitem [{\citenamefont {Aronovitz}(1988)}]{Aronovitz1988}%
  \BibitemOpen
  \bibfield  {author} {\bibinfo {author} {\bibfnamefont {J.~A.}\ \bibnamefont
  {Aronovitz}},\ }\bibfield  {title} {\enquote {\bibinfo {title} {Fluctuations
  of solid membranes},}\ }\href {\doibase 10.1103/PhysRevLett.60.2634}
  {\bibfield  {journal} {\bibinfo  {journal} {Phys. Rev. Lett.}\ }\textbf
  {\bibinfo {volume} {60}},\ \bibinfo {pages} {2634} (\bibinfo {year}
  {1988})}\BibitemShut {NoStop}%
\bibitem [{\citenamefont {Paczuski}\ \emph {et~al.}(1988)\citenamefont
  {Paczuski}, \citenamefont {Kardar},\ and\ \citenamefont
  {Nelson}}]{Paczuski1988}%
  \BibitemOpen
  \bibfield  {author} {\bibinfo {author} {\bibfnamefont {M.}~\bibnamefont
  {Paczuski}}, \bibinfo {author} {\bibfnamefont {M.}~\bibnamefont {Kardar}}, \
  and\ \bibinfo {author} {\bibfnamefont {D.~R.}\ \bibnamefont {Nelson}},\
  }\bibfield  {title} {\enquote {\bibinfo {title} {Landau theory of the
  crumpling transition},}\ }\href {\doibase 10.1103/PhysRevLett.60.2638}
  {\bibfield  {journal} {\bibinfo  {journal} {Phys. Rev. Lett.}\ }\textbf
  {\bibinfo {volume} {60}},\ \bibinfo {pages} {2638} (\bibinfo {year}
  {1988})}\BibitemShut {NoStop}%
\bibitem [{\citenamefont {David}\ and\ \citenamefont
  {Guitter}(1988)}]{David1988}%
  \BibitemOpen
  \bibfield  {author} {\bibinfo {author} {\bibfnamefont {F.}~\bibnamefont
  {David}}\ and\ \bibinfo {author} {\bibfnamefont {E.}~\bibnamefont
  {Guitter}},\ }\bibfield  {title} {\enquote {\bibinfo {title} {Crumpling
  transition in elastic membranes: Renormalization group treatment},}\ }\href
  {\doibase 10.1209/0295-5075/5/8/008} {\bibfield  {journal} {\bibinfo
  {journal} {Europhysics Lett. (EPL)}\ }\textbf {\bibinfo {volume} {5}},\
  \bibinfo {pages} {709} (\bibinfo {year} {1988})}\BibitemShut {NoStop}%
\bibitem [{\citenamefont {Aronovitz}\ \emph {et~al.}(1989)\citenamefont
  {Aronovitz}, \citenamefont {Golubovic},\ and\ \citenamefont
  {Lubensky}}]{Aronovitz1989}%
  \BibitemOpen
  \bibfield  {author} {\bibinfo {author} {\bibfnamefont {J.}~\bibnamefont
  {Aronovitz}}, \bibinfo {author} {\bibfnamefont {L.}~\bibnamefont
  {Golubovic}}, \ and\ \bibinfo {author} {\bibfnamefont {T.~C.}\ \bibnamefont
  {Lubensky}},\ }\bibfield  {title} {\enquote {\bibinfo {title} {Fluctuations
  and lower critical dimensions of crystalline membranes},}\ }\href {\doibase
  10.1051/jphys:01989005006060900} {\bibfield  {journal} {\bibinfo  {journal}
  {J. de Physique}\ }\textbf {\bibinfo {volume} {50}},\ \bibinfo {pages} {609}
  (\bibinfo {year} {1989})}\BibitemShut {NoStop}%
\bibitem [{\citenamefont {Guitter}\ \emph {et~al.}(1989)\citenamefont
  {Guitter}, \citenamefont {David}, \citenamefont {Leibler},\ and\
  \citenamefont {Peliti}}]{Guitter1989}%
  \BibitemOpen
  \bibfield  {author} {\bibinfo {author} {\bibfnamefont {E.}~\bibnamefont
  {Guitter}}, \bibinfo {author} {\bibfnamefont {F.}~\bibnamefont {David}},
  \bibinfo {author} {\bibfnamefont {S.}~\bibnamefont {Leibler}}, \ and\
  \bibinfo {author} {\bibfnamefont {L.}~\bibnamefont {Peliti}},\ }\bibfield
  {title} {\enquote {\bibinfo {title} {Thermodynamical behavior of polymerized
  membranes},}\ }\href {\doibase 10.1051/jphys:0198900500140178700} {\bibfield
  {journal} {\bibinfo  {journal} {J. de Physique}\ }\textbf {\bibinfo {volume}
  {50}},\ \bibinfo {pages} {1787} (\bibinfo {year} {1989})}\BibitemShut
  {NoStop}%
\bibitem [{\citenamefont {Le~Doussal}\ and\ \citenamefont
  {Radzihovsky}(1992)}]{Doussal1992}%
  \BibitemOpen
  \bibfield  {author} {\bibinfo {author} {\bibfnamefont {P.}~\bibnamefont
  {Le~Doussal}}\ and\ \bibinfo {author} {\bibfnamefont {L.}~\bibnamefont
  {Radzihovsky}},\ }\bibfield  {title} {\enquote {\bibinfo {title}
  {Self-consistent theory of polymerized membranes},}\ }\href {\doibase
  10.1103/PhysRevLett.69.1209} {\bibfield  {journal} {\bibinfo  {journal}
  {Phys. Rev. Lett.}\ }\textbf {\bibinfo {volume} {69}},\ \bibinfo {pages}
  {1209} (\bibinfo {year} {1992})}\BibitemShut {NoStop}%
\bibitem [{\citenamefont {Nelson}\ \emph {et~al.}(1989)\citenamefont {Nelson},
  \citenamefont {Piran},\ and\ \citenamefont {Weinberg}}]{BookNelson}%
  \BibitemOpen
  \bibinfo {editor} {\bibfnamefont {D.}~\bibnamefont {Nelson}}, \bibinfo
  {editor} {\bibfnamefont {T.}~\bibnamefont {Piran}}, \ and\ \bibinfo {editor}
  {\bibfnamefont {S.}~\bibnamefont {Weinberg}},\ eds.,\ \href@noop {} {\emph
  {\bibinfo {title} {Statistical Mechanics of Membranes and Surfaces}}}\
  (\bibinfo  {publisher} {World Scientific, Singapore},\ \bibinfo {year}
  {1989})\BibitemShut {NoStop}%
\bibitem [{\citenamefont {Le~Doussal}\ and\ \citenamefont
  {Radzihovsky}(2018)}]{Doussal2018}%
  \BibitemOpen
  \bibfield  {author} {\bibinfo {author} {\bibfnamefont {P.}~\bibnamefont
  {Le~Doussal}}\ and\ \bibinfo {author} {\bibfnamefont {L.}~\bibnamefont
  {Radzihovsky}},\ }\bibfield  {title} {\enquote {\bibinfo {title} {Anomalous
  elasticity, fluctuations and disorder in elastic membranes},}\ }\href
  {\doibase https://doi.org/10.1016/j.aop.2017.08.033} {\bibfield  {journal}
  {\bibinfo  {journal} {Ann. Phys. (N.Y.)}\ }\textbf {\bibinfo {volume}
  {392}},\ \bibinfo {pages} {340} (\bibinfo {year} {2018})}\BibitemShut
  {NoStop}%
\bibitem [{\citenamefont {Nicholl}\ \emph {et~al.}(2015)\citenamefont
  {Nicholl}, \citenamefont {Conley}, \citenamefont {Lavrik}, \citenamefont
  {Vlassiouk}, \citenamefont {Puzyrev}, \citenamefont {Sreenivas},
  \citenamefont {Pantelides},\ and\ \citenamefont {Bolotin}}]{Nicholl2015}%
  \BibitemOpen
  \bibfield  {author} {\bibinfo {author} {\bibfnamefont {R.~J.~T.}\
  \bibnamefont {Nicholl}}, \bibinfo {author} {\bibfnamefont {H.~J.}\
  \bibnamefont {Conley}}, \bibinfo {author} {\bibfnamefont {N.~V.}\
  \bibnamefont {Lavrik}}, \bibinfo {author} {\bibfnamefont {I.}~\bibnamefont
  {Vlassiouk}}, \bibinfo {author} {\bibfnamefont {Y.~S.}\ \bibnamefont
  {Puzyrev}}, \bibinfo {author} {\bibfnamefont {V.~P.}\ \bibnamefont
  {Sreenivas}}, \bibinfo {author} {\bibfnamefont {S.~T.}\ \bibnamefont
  {Pantelides}}, \ and\ \bibinfo {author} {\bibfnamefont {K.~I.}\ \bibnamefont
  {Bolotin}},\ }\bibfield  {title} {\enquote {\bibinfo {title} {The effect of
  intrinsic crumpling on the mechanics of free-standing graphene},}\ }\href
  {http://dx.doi.org/10.1038/ncomms9789} {\bibfield  {journal} {\bibinfo
  {journal} {Nat. Commun.}\ }\textbf {\bibinfo {volume} {6}},\ \bibinfo {pages}
  {9789} (\bibinfo {year} {2015})}\BibitemShut {NoStop}%
\bibitem [{\citenamefont {Gornyi}\ \emph {et~al.}(2016)\citenamefont {Gornyi},
  \citenamefont {Kachorovskii},\ and\ \citenamefont {Mirlin}}]{Gornyi2016}%
  \BibitemOpen
  \bibfield  {author} {\bibinfo {author} {\bibfnamefont {I.~V.}\ \bibnamefont
  {Gornyi}}, \bibinfo {author} {\bibfnamefont {V.~Yu.}\ \bibnamefont
  {Kachorovskii}}, \ and\ \bibinfo {author} {\bibfnamefont {A.~D.}\
  \bibnamefont {Mirlin}},\ }\bibfield  {title} {\enquote {\bibinfo {title}
  {Anomalous hooke's law in disordered graphene},}\ }\href {\doibase
  10.1088/2053-1583/4/1/011003} {\bibfield  {journal} {\bibinfo  {journal} {2D
  Materials}\ }\textbf {\bibinfo {volume} {4}},\ \bibinfo {pages} {011003}
  (\bibinfo {year} {2016})}\BibitemShut {NoStop}%
\bibitem [{\citenamefont {Ghorbanfekr-Kalashami}\ \emph
  {et~al.}(2017)\citenamefont {Ghorbanfekr-Kalashami}, \citenamefont {Vasu},
  \citenamefont {Nair}, \citenamefont {Peeters},\ and\ \citenamefont
  {Neek-Amal}}]{Ghorbanfekr2017}%
  \BibitemOpen
  \bibfield  {author} {\bibinfo {author} {\bibfnamefont {H.}~\bibnamefont
  {Ghorbanfekr-Kalashami}}, \bibinfo {author} {\bibfnamefont {K.~S.}\
  \bibnamefont {Vasu}}, \bibinfo {author} {\bibfnamefont {R.~R.}\ \bibnamefont
  {Nair}}, \bibinfo {author} {\bibfnamefont {Fran{\c c}ois~M.}\ \bibnamefont
  {Peeters}}, \ and\ \bibinfo {author} {\bibfnamefont {M.}~\bibnamefont
  {Neek-Amal}},\ }\bibfield  {title} {\enquote {\bibinfo {title} {Dependence of
  the shape of graphene nanobubbles on trapped substance},}\ }\href
  {http://dx.doi.org/10.1038/ncomms15844} {\bibfield  {journal} {\bibinfo
  {journal} {Nat. Commun.}\ }\textbf {\bibinfo {volume} {8}} (\bibinfo {year}
  {2017})}\BibitemShut {NoStop}%
\bibitem [{\citenamefont {Zhang}\ and\ \citenamefont
  {Arroyo}(2017)}]{Zhang2017}%
  \BibitemOpen
  \bibfield  {author} {\bibinfo {author} {\bibfnamefont {K.}~\bibnamefont
  {Zhang}}\ and\ \bibinfo {author} {\bibfnamefont {M.}~\bibnamefont {Arroyo}},\
  }\bibfield  {title} {\enquote {\bibinfo {title} {Coexistence of wrinkles and
  blisters in supported graphene},}\ }\href
  {http://www.sciencedirect.com/science/article/pii/S2352431616301845}
  {\bibfield  {journal} {\bibinfo  {journal} {Extreme Mechanics Lett.}\
  }\textbf {\bibinfo {volume} {14}},\ \bibinfo {pages} {23} (\bibinfo {year}
  {2017})}\BibitemShut {NoStop}%
\bibitem [{\citenamefont {Delfani}(2018)}]{Delfanu2018}%
  \BibitemOpen
  \bibfield  {author} {\bibinfo {author} {\bibfnamefont {M.R.}\ \bibnamefont
  {Delfani}},\ }\bibfield  {title} {\enquote {\bibinfo {title} {Nonlinear
  elasticity of monolayer hexagonal crystals: Theory and application to
  circular bulge test},}\ }\href
  {http://www.sciencedirect.com/science/article/pii/S0997753816304594}
  {\bibfield  {journal} {\bibinfo  {journal} {European Journal of Mechanics -
  A/Solids}\ }\textbf {\bibinfo {volume} {68}},\ \bibinfo {pages} {117}
  (\bibinfo {year} {2018})}\BibitemShut {NoStop}%
\bibitem [{\citenamefont {Sanchez}\ \emph {et~al.}(2018)\citenamefont
  {Sanchez}, \citenamefont {Dai}, \citenamefont {Wang}, \citenamefont
  {Brennan}, \citenamefont {Huang},\ and\ \citenamefont {Lu}}]{Sanchez2018}%
  \BibitemOpen
  \bibfield  {author} {\bibinfo {author} {\bibfnamefont {D.~A.}\ \bibnamefont
  {Sanchez}}, \bibinfo {author} {\bibfnamefont {Z.}~\bibnamefont {Dai}},
  \bibinfo {author} {\bibfnamefont {A.}~\bibnamefont {Wang}, \bibfnamefont
  {Z.and Cantu-Chavez}}, \bibinfo {author} {\bibfnamefont {C.~J.}\ \bibnamefont
  {Brennan}}, \bibinfo {author} {\bibfnamefont {R.}~\bibnamefont {Huang}}, \
  and\ \bibinfo {author} {\bibfnamefont {N.}~\bibnamefont {Lu}},\ }\bibfield
  {title} {\enquote {\bibinfo {title} {Mechanics of spontaneously formed
  nanoblisters trapped by transferred {2D} crystals},}\ }\href@noop {}
  {\bibfield  {journal} {\bibinfo  {journal} {PNAS}\ }\textbf {\bibinfo
  {volume} {115}},\ \bibinfo {pages} {7884} (\bibinfo {year}
  {2018})}\BibitemShut {NoStop}%
\bibitem [{\citenamefont {Landau}\ and\ \citenamefont
  {Lifshitz}(2012)}]{Landau}%
  \BibitemOpen
  \bibfield  {author} {\bibinfo {author} {\bibfnamefont {L.D.}\ \bibnamefont
  {Landau}}\ and\ \bibinfo {author} {\bibfnamefont {E.M.}\ \bibnamefont
  {Lifshitz}},\ }\href@noop {} {\emph {\bibinfo {title} {Theory of
  Elasticity}}}\ (\bibinfo  {publisher} {Butterworth-Heinemann},\ \bibinfo
  {year} {2012})\BibitemShut {NoStop}%
\bibitem [{\citenamefont {Bowick}\ \emph {et~al.}(1996)\citenamefont {Bowick},
  \citenamefont {Catterall}, \citenamefont {Falcioni}, \citenamefont
  {Thorleifsson},\ and\ \citenamefont {Anagnostopoulos}}]{Bowick1996}%
  \BibitemOpen
  \bibfield  {author} {\bibinfo {author} {\bibfnamefont {M.~J.}\ \bibnamefont
  {Bowick}}, \bibinfo {author} {\bibfnamefont {S.~M.}\ \bibnamefont
  {Catterall}}, \bibinfo {author} {\bibfnamefont {M.}~\bibnamefont {Falcioni}},
  \bibinfo {author} {\bibfnamefont {G.}~\bibnamefont {Thorleifsson}}, \ and\
  \bibinfo {author} {\bibfnamefont {K.~N.}\ \bibnamefont {Anagnostopoulos}},\
  }\bibfield  {title} {\enquote {\bibinfo {title} {The flat phase of
  crystalline membranes},}\ }\href {http://dx.doi.org/10.1051/jp1:1996139}
  {\bibfield  {journal} {\bibinfo  {journal} {J. de Physique I}\ }\textbf
  {\bibinfo {volume} {6}},\ \bibinfo {pages} {1321} (\bibinfo {year}
  {1996})}\BibitemShut {NoStop}%
\bibitem [{\citenamefont {Tr{\"o}ster}(2013)}]{Troster2013}%
  \BibitemOpen
  \bibfield  {author} {\bibinfo {author} {\bibfnamefont {A.}~\bibnamefont
  {Tr{\"o}ster}},\ }\bibfield  {title} {\enquote {\bibinfo {title} {Fourier
  monte carlo simulation of crystalline membranes in the flat phase},}\ }\href
  {http://dx.doi.org/10.1088/1742-6596/454/1/012032} {\bibfield  {journal}
  {\bibinfo  {journal} {J. Phys.: Conf. Series}\ }\textbf {\bibinfo {volume}
  {454}},\ \bibinfo {pages} {012032} (\bibinfo {year} {2013})}\BibitemShut
  {NoStop}%
\bibitem [{\citenamefont {Rold{\'a}n}\ \emph {et~al.}(2011)\citenamefont
  {Rold{\'a}n}, \citenamefont {Fasolino}, \citenamefont {Zakharchenko},\ and\
  \citenamefont {Katsnelson}}]{Roldan2011}%
  \BibitemOpen
  \bibfield  {author} {\bibinfo {author} {\bibfnamefont {R.}~\bibnamefont
  {Rold{\'a}n}}, \bibinfo {author} {\bibfnamefont {A.}~\bibnamefont
  {Fasolino}}, \bibinfo {author} {\bibfnamefont {K.~V.}\ \bibnamefont
  {Zakharchenko}}, \ and\ \bibinfo {author} {\bibfnamefont {M.~I.}\
  \bibnamefont {Katsnelson}},\ }\bibfield  {title} {\enquote {\bibinfo {title}
  {Suppression of anharmonicities in crystalline membranes by external
  strain},}\ }\href {\doibase 10.1103/PhysRevB.83.174104} {\bibfield  {journal}
  {\bibinfo  {journal} {Phys. Rev. B}\ }\textbf {\bibinfo {volume} {83}},\
  \bibinfo {pages} {174104} (\bibinfo {year} {2011})}\BibitemShut {NoStop}%
\bibitem [{\citenamefont {Ko{\v s}mrlj}\ and\ \citenamefont
  {Nelson}(2016)}]{Kosmrlj2016}%
  \BibitemOpen
  \bibfield  {author} {\bibinfo {author} {\bibfnamefont {A.}~\bibnamefont
  {Ko{\v s}mrlj}}\ and\ \bibinfo {author} {\bibfnamefont {D.~R.}\ \bibnamefont
  {Nelson}},\ }\bibfield  {title} {\enquote {\bibinfo {title} {Response of
  thermalized ribbons to pulling and bending},}\ }\href {\doibase
  10.1103/PhysRevB.93.125431} {\bibfield  {journal} {\bibinfo  {journal} {Phys.
  Rev. B}\ }\textbf {\bibinfo {volume} {93}},\ \bibinfo {pages} {125431}
  (\bibinfo {year} {2016})}\BibitemShut {NoStop}%
\bibitem [{\citenamefont {Burmistrov}\ \emph {et~al.}(2016)\citenamefont
  {Burmistrov}, \citenamefont {Gornyi}, \citenamefont {Kachorovskii},
  \citenamefont {Katsnelson},\ and\ \citenamefont {Mirlin}}]{Burmistrov2016}%
  \BibitemOpen
  \bibfield  {author} {\bibinfo {author} {\bibfnamefont {I.~S.}\ \bibnamefont
  {Burmistrov}}, \bibinfo {author} {\bibfnamefont {I.~V.}\ \bibnamefont
  {Gornyi}}, \bibinfo {author} {\bibfnamefont {V.~Yu.}\ \bibnamefont
  {Kachorovskii}}, \bibinfo {author} {\bibfnamefont {M.~I.}\ \bibnamefont
  {Katsnelson}}, \ and\ \bibinfo {author} {\bibfnamefont {A.~D.}\ \bibnamefont
  {Mirlin}},\ }\bibfield  {title} {\enquote {\bibinfo {title} {Quantum
  elasticity of graphene: Thermal expansion coefficient and specific heat},}\
  }\href {\doibase 10.1103/PhysRevB.94.195430} {\bibfield  {journal} {\bibinfo
  {journal} {Phys. Rev. B}\ }\textbf {\bibinfo {volume} {94}},\ \bibinfo
  {pages} {195430} (\bibinfo {year} {2016})}\BibitemShut {NoStop}%
\bibitem [{Note1()}]{Note1}%
  \BibitemOpen
  \bibinfo {note} {Here we neglect weak logarithmic dependence on ${R}$ of the
  bending rigidity for ${R\gg R_\sigma }$ (cf. {R}ef. \cite {Kosmrlj2016}).
  {S}uch logarithmic corrections are beyond accuracy of our
  estimates.}\BibitemShut {Stop}%
\bibitem [{\citenamefont {Paulose}\ \emph {et~al.}(2012)\citenamefont
  {Paulose}, \citenamefont {Vliegenthart}, \citenamefont {Gompper},\ and\
  \citenamefont {Nelson}}]{Paulose2012}%
  \BibitemOpen
  \bibfield  {author} {\bibinfo {author} {\bibfnamefont {J.}~\bibnamefont
  {Paulose}}, \bibinfo {author} {\bibfnamefont {G.~A.}\ \bibnamefont
  {Vliegenthart}}, \bibinfo {author} {\bibfnamefont {G.}~\bibnamefont
  {Gompper}}, \ and\ \bibinfo {author} {\bibfnamefont {D.~R.}\ \bibnamefont
  {Nelson}},\ }\bibfield  {title} {\enquote {\bibinfo {title} {Fluctuating
  shells under pressure},}\ }\href@noop {} {\bibfield  {journal} {\bibinfo
  {journal} {PNAS}\ }\textbf {\bibinfo {volume} {109}},\ \bibinfo {pages}
  {19551} (\bibinfo {year} {2012})}\BibitemShut {NoStop}%
\bibitem [{SM()}]{SM}%
  \BibitemOpen
  \href@noop {} {}\bibinfo {note} {See Supplemental Material}\BibitemShut
  {NoStop}%
\bibitem [{\citenamefont {Zhilyaev}\ \emph {et~al.}(2019)\citenamefont
  {Zhilyaev}, \citenamefont {Iakovlev},\ and\ \citenamefont
  {Akhatov}}]{Zhilyaev2019}%
  \BibitemOpen
  \bibfield  {author} {\bibinfo {author} {\bibfnamefont {P.}~\bibnamefont
  {Zhilyaev}}, \bibinfo {author} {\bibfnamefont {E.}~\bibnamefont {Iakovlev}},
  \ and\ \bibinfo {author} {\bibfnamefont {I.}~\bibnamefont {Akhatov}},\
  }\bibfield  {title} {\enquote {\bibinfo {title} {Liquid--gas phase transition
  of ar inside graphene nanobubbles on the graphite substrate},}\ }\href
  {\doibase 10.1088/1361-6528/ab061f} {\bibfield  {journal} {\bibinfo
  {journal} {Nanotechnology}\ }\textbf {\bibinfo {volume} {30}},\ \bibinfo
  {pages} {215701} (\bibinfo {year} {2019})}\BibitemShut {NoStop}%
\bibitem [{\citenamefont {Ko{\v s}mrlj}\ and\ \citenamefont
  {Nelson}(2017)}]{Kosmrlj2017}%
  \BibitemOpen
  \bibfield  {author} {\bibinfo {author} {\bibfnamefont {A.}~\bibnamefont
  {Ko{\v s}mrlj}}\ and\ \bibinfo {author} {\bibfnamefont {D.~R.}\ \bibnamefont
  {Nelson}},\ }\bibfield  {title} {\enquote {\bibinfo {title} {Statistical
  mechanics of thin spherical shells},}\ }\href {\doibase
  10.1103/PhysRevX.7.011002} {\bibfield  {journal} {\bibinfo  {journal} {Phys.
  Rev. X}\ }\textbf {\bibinfo {volume} {7}},\ \bibinfo {pages} {011002}
  (\bibinfo {year} {2017})}\BibitemShut {NoStop}%
\end{thebibliography}%



\section*{Supplementary material}

# Supplemental Material for "The effect of anomalous elasticity on the profile of bubbles in van der Waals heterostructures"

A. A. Lyublinskaya,[1] S. S. Babkin,[1] and I. S. Burmistrov[2,3]

[1]*Moscow Institute of Physics and Technology, 141700, Dolgoprudnyi, Moscow Region, Russia*
[2]*L. D. Landau Institute for Theoretical Physics, Semenova 1-a, 142432, Chernogolovka, Russia*
[3]*Laboratory for Condensed Matter Physics, National Research University Higher School of Economics, 101000 Moscow, Russia*

This material contains (i) the accurate analytical calculation of the maximal height of the bubble in the bending dominating regime; (ii) the small volume regime of the liquid bubble; and (iii) the estimates demonstrating the absence of buckling transition of the bubble; (iii) the estimate for thermodynamic fluctuations of the height of the bubble.

## S.I. THE MAXIMAL HEIGHT OF THE BUBBLE IN THE BENDING DOMINATING REGIME

Let us introduce the normalized eigenfunctions of the Laplace operator on the disk $r \leqslant R$ satisfying zero boundary conditions at $r = R$:

$$\Delta \phi_n(\mathbf{r}) = -\frac{\zeta_n^2}{R^2} \phi_n(\mathbf{r}), \qquad \phi_n(\mathbf{r}) = \frac{1}{\sqrt{\pi} R |J_1(\zeta_n)|} J_0\left(\frac{\zeta_n r}{R}\right). \tag{S1}$$

Here $\zeta_n$ with $n = 1, 2, \ldots$ are the zeroes of the zeroth order Bessel function $J_0(x)$. Then, one can write the renormalized bending energy as follows

$$E_{\text{bend}} = \frac{1}{2} \int d^2\mathbf{r}\, d^2\mathbf{r}'\, \Delta h(\mathbf{r}) \hat{\varkappa}(\mathbf{r} - \mathbf{r}') \Delta h(\mathbf{r}'), \tag{S2}$$

where the integral operator $\hat{\varkappa}$ is defined via its action on the eigenfunctions of the Laplace operator:

$$\int d^2\mathbf{r}'\, \hat{\varkappa}(\mathbf{r} - \mathbf{r}') \phi_n(\mathbf{r}') = \zeta_n^{-\eta} \varkappa(R) \phi_n(\mathbf{r}). \tag{S3}$$

Here we remind $\varkappa(R) = \varkappa_0 (R/R_*)^\eta$.

Now let us expand the height of the bubble into the series: $h(\mathbf{r}) = \sum_n \alpha_n \phi_n(\mathbf{r})$. This expansion automatically satisfies the boundary condition $h(r = R) = 0$. The knowledge of the coefficients $\alpha_n$ allows one to find the maximal height of the bubble: $H = \sum_n \alpha_n / [\sqrt{\pi} R |J_1(\zeta_n)|]$. In terms of the coefficients $\alpha_n$ the total energy acquires the following form:

$$E_{\text{bend}} + E_{\text{b}} + E_{\text{vdW}} = \frac{\varkappa(R)}{2R^4} \sum_n \zeta_n^{4-\eta} \alpha_n^2 - 2\sqrt{\pi} P R \sum_n \frac{\operatorname{sgn}[J_1(\zeta_n)]}{\zeta_n} \alpha_n + \pi \gamma R^2. \tag{S4}$$

Its minimization over $\alpha_n$ yeilds

$$\alpha_n = 2\sqrt{\pi} \frac{P R^5}{\varkappa(R)} \frac{\operatorname{sgn}[J_1(\zeta_n)]}{\zeta_n^{5-\eta}}. \tag{S5}$$

Hence, for the total energy we obtain

$$E_{\text{bend}} + E_{\text{b}} + E_{\text{vdW}} = -2\pi \frac{P^2 R^6}{\varkappa(R)} \sum_n \zeta_n^{-6+\eta} + \pi \gamma R^2. \tag{S6}$$

Minimization with respect to $R$ determines the pressure inside the bubble:

$$P = c_4' \frac{\sqrt{\gamma \varkappa(R)}}{R^2}, \qquad c_4' = \left[(6-\eta) \sum_n \zeta_n^{-6+\eta}\right]^{-1/2} \approx 9.72 \tag{S7}$$

Using this result for the pressure, we find the final expression for the maximal height:

$$H = 2\frac{P R^4}{\varkappa(R)} \sum_n \frac{\zeta_n^{-5+\eta}}{J_1(\zeta_n)} = c_4 R^2 \left(\frac{\gamma}{\varkappa(R)}\right)^{1/2}, \qquad c_4 = 2 c_4' \sum_n \frac{\zeta_n^{-5+\eta}}{J_1(\zeta_n)} \approx 0.90. \tag{S8}$$

The constants $c_3$ and $c_3'$ relevant for the low temperature regime $R \ll R_*$ are obtained from the results above by setting $\eta = 0$. Then one finds $c_3 \approx 0.65$ and $c_3' \approx 13.86$.

## S.II. THE LIQUID BUBBLE OF A SMALL VOLUME $V \ll V_\gamma$

In the case of the bubble with small volume of liquid, $V \ll V_\gamma$ the temperature dependence of the maximal height is determined by the bending dominated regime. At low temperatures from Eq. 6 of the main text we obtain that the maximal height of the bubble is independent of temperature:

$$H \sim a(V/V_\gamma)^{1/2}, \qquad T \ll T_\gamma(V/V_\gamma)^{-1/2}. \tag{S9}$$

At high temperatures, as it follows from Eq. 10 of the main text, the maximal height of the bubble starts to decrease with the temperature:

$$H \sim a(V/V_\gamma)^{\frac{4-\eta}{8-\eta}}(T/T_\gamma)^{-\frac{\eta}{8-\eta}}, \qquad T_\gamma(V/V_\gamma)^{-\frac{1}{2}} \ll T. \tag{S10}$$

We note that the power law decay is controlled by the bending rigidity exponent $\eta$. The decay of $H$ with increase of $T$ implies the negative linear thermal expansion coefficient $\alpha_H = (1/H)dH/dT$. Although interesting on its own, the bubbles with a liquid of small volume $V \ll V_\gamma$ can hardly be detected since as it follows from Eqs. S9 and S10 the maximal height of the bubble is smaller than $a$.

## S.III. ABSENCE OF THE BUCKLING INSTABILITY OF A SPHERICAL BUBBLE

In the presence of non-zero curvature radius $R_0$ of the membrane flexural phonons acquire a true mass equal to $Y_0/R_0^2$ [S1]. This modification of the spectrum of flexural phonons results in negative renormalization of the tension. Perturbatively, a change of the tension $\delta\sigma$ can be estimated as [S1]:

$$\frac{\delta\sigma}{\sigma} \sim -\frac{TY_0^2}{\sigma R_0^2}\int \frac{d^2\boldsymbol{k}}{(2\pi)^2}\frac{k^2}{(\varkappa_0 k^4 + \sigma k^2 + Y_0/R_0^2)^2}. \tag{S11}$$

We emphasize that here the integral over momentum $k$ has infrared cut off due to the radius of the bubble $R$: $k \gtrsim 1/R$.

Eq. S11 can be directly applied to the regime of low temperatures $T \ll T_\gamma$. Then, using Eq. 6 of the main text, we find

$$\frac{\delta\sigma}{\sigma} \sim -\frac{T}{\gamma R^2}\ln\frac{R^2\sqrt{\gamma Y_0}}{\varkappa_0}. \tag{S12}$$

Since the low temperature theory is applicable for $R \gg a(Y_0/\gamma)^{1/4}$, we find that $|\delta\sigma|/\sigma \ll (T/T_\gamma)^2 \ll 1$.

Now let us consider the regime of high temperatures $T \gg T_\gamma$. For $R \ll R_*$ we can use Eq. S11. Then, using Eq. 9 of the main text, we obtain ($x = kR$)

$$\frac{\delta\sigma}{\sigma} \sim -\left(\frac{T_\gamma}{T}\right)^2\left(\frac{R}{R_*}\right)^6\int_{\sim 1} dx \frac{x^3}{(x^4 + x^2 + \gamma Y_0 R^4/\varkappa_0^2)^2}. \tag{S13}$$

Since the parameter $\gamma Y_0 R^4/\varkappa_0^2)^2 \sim (T_\gamma/T)^2(R/R_*)^4 \ll 1$, integrating over $x$, we find

$$\frac{\delta\sigma}{\sigma} \sim -\left(\frac{T_\gamma}{T}\right)^2\left(\frac{R}{R_*}\right)^6. \tag{S14}$$

Therefore we find that $|\delta\sigma|/\sigma \ll (T_\gamma/T)^2 \ll 1$ for $R \ll R_*$. In the case of intermediate radius $R_* \ll R \ll R_\sigma = R_*T/T_\gamma$, one needs to modify Eq. S11 in order to include renormalization of the bending rigidity and Young's modulus:

$$\frac{\delta\sigma}{\sigma} \sim -\frac{TY(R)^2}{\sigma R_0^2}\int \frac{d^2\boldsymbol{k}}{(2\pi)^2}\frac{k^2}{(\varkappa(R)k^4 + \sigma k^2 + Y(R)/R_0^2)^2}. \tag{S15}$$

Using Eq. 10 of the main text, we obtain

$$\frac{\delta\sigma}{\sigma} \sim -\left(\frac{R}{R_\sigma}\right)^2\int_{\sim 1} dx \frac{x^3}{(x^4 + x^2 + \gamma Y(R)R^4/\varkappa^2(R))^2}. \tag{S16}$$

Since $\gamma Y(R)R^4/\varkappa^2(R) \sim (R/R_\sigma)^2 \ll 1$, we obtain

$$\frac{|\delta\sigma|}{\sigma} \sim \left(\frac{R}{R_\sigma}\right)^2 \ll 1, \qquad R_* \ll R \ll R_\sigma. \tag{S17}$$

Finally, we consider the case $R \gg R_\sigma$. Here we can use Eq. S12 with $\varkappa_0$ and $Y_0$ substituted by $\varkappa(R_\sigma)$ and $Y(R_\sigma)$, respectively. Then, we find

$$\frac{|\delta\sigma|}{\sigma} \sim \left(\frac{R_\sigma}{R}\right)^2 \ln\frac{R}{R_\sigma} \ll 1, \qquad R \gg R_\sigma. \tag{S18}$$

The above results demonstrates that for both low and high temperature regimes, the renormalization of the tension due to the finite curvature is weak and cannot lead to the buckling instability. The only exception is the bubbles with the radius $R \sim R_\sigma$ for which our estimates give $|\delta\sigma|/\sigma \sim 1$. This implies that for $R \sim R_\sigma$ the buckling instability could occur in principle. In order to resolve this issue one needs to perform more accurate RG scheme similar to the one of Ref. [S2]. However, since for $R \ll R_\sigma$ and $R \gg R_\sigma$ the buckling instability is absent the scenario with instability for $R \sim R_\sigma$ seems to be unlikely.

## S.IV. THE THERMODYNAMIC FLUCTUATIONS OF THE HEIGHT OF THE BUBBLE

Let us estimate the thermodynamic fluctuations of the height of the bubble in the bending dominated regime, $R \ll R_* T/T_\gamma$. Using Eq. 4 of the main text, we find

$$\langle(\delta H)^2\rangle \sim T \int \frac{d^2\boldsymbol{k}}{(2\pi)^2} \frac{1}{\varkappa(R)k^4}. \tag{S19}$$

As above the integral over momentum $k$ has infrared cut off due to the radius of the bubble $R$: $k \gtrsim 1/R$. Hence we find that the dispersion of height fluctuations is given as follows

$$\langle(\delta H)^2\rangle \sim TR^2/\varkappa(R). \tag{S20}$$

Then for $R \ll R_* T/T_\gamma$ we obtain the following estimate:

$$\frac{\sqrt{\langle(\delta H)^2\rangle}}{H} \sim \frac{T}{T_\gamma}\frac{R_*}{R} \gg 1. \tag{S21}$$

This inequality implies that there are large thermodynamic fluctuations of the height of the bubble in the bending dominated regime.

For the tension dominated regime $R \gg R_\sigma = R_* T/T_\gamma$ one needs to take into account the tension induced by the pressure:

$$\langle(\delta H)^2\rangle \sim T \int \frac{d^2\boldsymbol{k}}{(2\pi)^2} \frac{1}{\varkappa(R_\sigma)k^4 + \sigma_P k^2} \sim \frac{TR_\sigma^2}{\varkappa(R_\sigma)} \ln\frac{R}{R_\sigma}. \tag{S22}$$

Then we obtain the following estimate:

$$\frac{\sqrt{\langle(\delta H)^2\rangle}}{H} \sim \frac{T}{T_\gamma}\frac{R_*}{R} \ln\frac{R}{R_\sigma} \ll 1. \tag{S23}$$

Therefore the thermodynamic fluctuations of the height of the bubble in the tension dominated regime are strongly suppressed.

[S1] J. Paulose, G. A. Vliegenthart, G. Gompper, and D. R. Nelson, *Fluctuating shells under pressure*, PNAS **109**, 19551 (2012).
[S2] A. Kosmrlj and D. R. Nelson, *Statistical mechanics of thin spherical shells*, Phys. Rev. X **7**, 011002 (2017).



\end{document}